\documentstyle[aps,epsf,epsfig]{revtex}

\begin{document}
\author{D. Foerster, CPTMB, Universit\'{e} de Bordeaux I}
\address{351, cours de la Liberation, F - 33405 Talence Cedex}
\title{A planar extrapolation of the correlation problem that permits pairing}
\maketitle

\begin{abstract}
It was observed previously that the $SU(N)$ extension of the Hubbard model
is dominated by planar diagrams at large $N$, but the possibility of
superconducting pairing got lost in this extrapolation. To allow for
pairing, we replace $SU(N)$ by $U(N,q)$, the unitary group in a vector space
of quaternions. At the level of the free energy, the difference between the $%
SU(N)$ and $U(N,q)$ extrapolations appears only to first nonleading order in 
$N$.

\end{abstract}

\section{Introduction and Motivation}

The Hubbard model \cite{Hubbard} belongs to a class of models describing
strongly correlated electrons in systems such as, for example, high $T_{c}$
ceramics, transition metal oxides and organic conductors\cite{Fulde}.
However, in spite of much progress made over the last $\sim 40$ years, it is
still difficult to determine the low temperature properties of such model
systems as a function of their chemical input parameters.

Recently it was noticed \cite{DF} that an $SU(N)$ deformation of such models
is dominated, for $N\gg 1$, by planar diagrams in the sense of 't Hooft \cite
{Hooft} and the sum of the leading planar diagrams turned out to be very
similar to the ''Fluctuation Exchange Approximation (FLEX)'' pioneered by Bickers and
Scalapino \cite{Scalapino}. However, unlike the FLEX approximation, the new
method permits a systematic construction of the generating functional in
powers of $1/N$.

To exploit this new perspective on the correlation problem, we should study
the evolution of the properties of a given model as one moves from $N=\infty 
$ to $N=2$. But precisely such a study is impossible in the $SU(N)$
extrapolation of the Hubbard model since for $N>2$ it contains no analog of
the superconducting order parameter $<\psi _{\alpha }(x)\varepsilon _{\alpha
\beta }\psi _{\beta }(y)>$. The very feature that constituted one of the
motivations of studying Hubbard like models in the first place got lost in
this extrapolation.

Some time ago it was recognized\cite{Sachdev}, that $U(N,q)$, the unitary
group acting in a vector space of quaternions, provides an interesting way
of implementing the large $N$ limit because it generalizes the antisymmetric
tensor $\varepsilon _{\alpha \beta }$ in a natural way. Because of $%
U(1,q)=SU(2)$, an extrapolation via $U(N,q)$ with $N=1,2,..$ is just as good
an extrapolation as $SU(N)$ with $N=2,3..$ Results of what probably amounted
to an $U(N,q)$ extrapolation of the $tJ$ model were given recently \cite
{stripes}, but without any calculational details.

We will show below that there is a unique $U(N,q)$ extrapolation of the
Hubbard model that permits pairing and, simultaneously, a topological
expansion in which the planar diagrams dominate for $N\gg 1$.

\section{$U(N,q)$ invariant interactions between bosons and fermions}

It is reasonable to begin our discussion by recalling the definition of the
group $U(N,q)$ \cite{Chevalley}. It acts in a vector space of quaternions: 
\begin{eqnarray}
\psi _{a} &=&\psi _{a,0}+i\sum_{r=1..3}\psi _{a,r}\sigma ^{r}\text{ \ \ \ \
and\ \ \ \ \ \ }\psi _{a}^{\ast }=\psi _{a,0}-i\sum_{r=1..3}\psi
_{a,r}\sigma ^{r}  \label{quaternions} \\
&&\psi _{a\mu }\text{ real for }a=1..N\text{, }\mu =0..3\text{, \ \ \ }%
\overrightarrow{\sigma }=\text{Pauli matrices}  \nonumber
\end{eqnarray}
The $\psi _{a\mu }$ are real because we wish to think of the matrices $%
i\sigma ^{r}$, $r=1..3$ as square roots of $-1$. The transformation matrix $%
g $ is also quaternion valued, $g=g^{0}+i\overrightarrow{g}\overrightarrow{%
\sigma }$ with $g_{\mu }$, $\mu =0..3$ real in order for the image to be a
quaternion. As a further constraint, it must leave a quaternion valued
length invariant: 
\begin{equation}
\sum_{a=1..N}\psi _{a}^{\ast }\psi _{a}=\sum_{a=1..N}(\psi ^{\ast
}g^{+})_{a}(g\psi )_{a}\rightarrow g^{+}g=1  \label{unitary}
\end{equation}
The properties of the Pauli matrices $\overrightarrow{\sigma }$ imply $%
g^{T}=\sigma ^{2}\left( g^{0}-i\overrightarrow{g}\overrightarrow{\sigma }%
\right) \sigma ^{2}$ and $g^{+}=g^{0}-i\overrightarrow{g}\overrightarrow{%
\sigma }$ from which we deduce $g^{T}=\sigma ^{2}g^{+}\sigma ^{2}$. In
conjunction with $g^{+}g=1$ this latter relation implies $g^{T}\sigma
^{2}g=\sigma ^{2}$. The relations 
\begin{equation}
g^{+}g=1\text{ and }g^{T}\sigma ^{2}g=\sigma ^{2}  \label{TwoInvariants}
\end{equation}
can be restated by saying that $U(N,q)$ has two invariant tensors,$\ \delta
_{ab}\sigma _{\alpha \beta }^{2}$ and $\delta _{ab}\delta _{\alpha \beta }$.

The quaternions $\psi _{a}\rightarrow $ $\left( \psi _{a}\right) _{\sigma
\tau }$ of eq(\ref{quaternions}) are not all independent because they can be
parametrized by only 4 real quantities for each value of $a$. To get rid of
the redundant quantities, while keeping the attractive transformation
properties of the quaternions, we define ''conventional'' complex electron
operators 
\begin{equation}
\psi _{a\sigma }=\sum_{\tau =1,2}\left( \psi _{a}\right) _{\sigma \tau }\xi
_{\tau }\text{ \ \ , \ \ \ \ \ }\xi =\left( 
\begin{array}{c}
1 \\ 
0
\end{array}
\right) 
\end{equation}
that have only one spinor index. 

A nonvanishing average such as 
\begin{equation}
<\sum \psi _{a\sigma }(x)\sigma _{\sigma \tau }^{2}\psi _{a\tau }(y)>\neq 0
\end{equation}
breaks charge but conserves $U(N,q)$ \ and pairing is therefore possible in
an $U(N,q)$ extension of the Hubbard model.

Following the strategy described in \cite{DF} we will introduce auxiliary
boson fields via a Hubbard Stratonovich transformation. For this we need (i)
an invariant quadratic form in the auxiliary bosons and (ii) an invariant
interaction between the auxiliary bosons and some fermion bilinears. Both
requirements are met if the auxiliary fields are infinitesimal generators of 
$U(N,q)$, so we consider them now. Writing finite transformations as $g=\exp
iz$ , \ \ $z=iz_{0}+\overrightarrow{z}\overrightarrow{\sigma }$ with $z_{\mu
}$ complex, for the moment, and using $g^{+}g=1$ and $g^{T}\sigma
^{2}g=\sigma ^{2}$ we find 
\begin{equation}
z^{+}=z\text{ and }z^{T}\sigma ^{2}+\sigma ^{2}z=0  \label{invariants}
\end{equation}
The first of these equations implies $\left\{ z_{0}^{+}=-z_{0},%
\overrightarrow{z}^{+}=\overrightarrow{z}\right\} $ while the second means $%
\left\{ z_{0}^{T}=-z_{0},\overrightarrow{z}^{T}=\overrightarrow{z}\right\} $%
. Both taken together imply that $z_{\mu }$ is real so that $iz$ is a
quaternion valued matrix (and therefore $g=\exp iz$ \ also a quaternion).
The preceding argument shows that $U(N,q)$ depends on a total of $\frac{1}{2}%
N(N-1)+\frac{3}{2}N(N+1)=2N^{2}+N$ independent parameters. In particular,
for $N=1$, there are $3$ generators, the antisymmetric matrix $z_{0}$ is
absent and from the explicit parametrization $g=\exp i\overrightarrow{z}%
\overrightarrow{\sigma }$ we conclude that $U(1,q)=SU(2)$. This conclusion
is also very natural in view of the definition of the group according to eq(%
\ref{unitary}). The quadratic form in the Lie algebra 
\begin{eqnarray}
||z||^{2} &=&\frac{1}{2}Tr(zz^{+})=\frac{1}{2}Tr\left( iz_{0}+%
\overrightarrow{z}\overrightarrow{\sigma }\right) (-iz_{0}^{+}+%
\overrightarrow{z}^{+}\overrightarrow{\sigma })  \label{quadratic} \\
&=&Tr(z_{0}z_{0}^{+}+\overrightarrow{z}\overrightarrow{z}^{+})=\sum
z_{ab}^{\mu }z_{ab}^{\mu }  \nonumber
\end{eqnarray}
is positive and invariant and an invariant coupling between fermions and
bosons is given by $\psi ^{+}\left( iz^{0}+\overrightarrow{z}\overrightarrow{%
\sigma }\right) \psi $. Because of its symmetry properties, $z_{\mu }$
couples only to roughly half the bilinears in $\psi ^{+}$and $\psi $ in this
expression: 
\begin{eqnarray}
\psi ^{+}\left( iz_{0}+\overrightarrow{z}\overrightarrow{\sigma }\right)
\psi  &=&\sum z_{ab}^{0}\frac{i}{2}\left[ \psi _{a\alpha }^{+}\psi _{b\alpha
}-\left( a\leftrightarrow b\right) \right] +\overrightarrow{z}_{ab}\frac{1}{2%
}\left[ \psi _{a\alpha }^{+}\overrightarrow{\sigma }_{\alpha \beta }\psi
_{b\beta }+\left( a\leftrightarrow b\right) \right] =\sum z_{ab}^{\mu
}J_{ab}^{\mu }  \label{current} \\
\text{ with }J_{ab}^{\mu } &=&\frac{1}{2}\left[ \psi _{a\alpha }^{+}\sigma
_{\alpha \beta }^{\mu }\psi _{b\beta }+h.c.\right] \text{ and }\sigma ^{\mu
}=\left\{ i,\overrightarrow{\sigma }\right\}   \nonumber
\end{eqnarray}
Indeed, symmetrization with respect to Hermitian conjugation amounts to
symmetrization with respect to the indices $a,b$.

\section{Hamiltonian and partition function}

The infinitesimal generators $z$ can be converted into dynamical bosons $%
z(\tau )$ with the help of a Hubbard Stratonovich transformation: 
\begin{eqnarray}
const &=&\int Dz\exp -\int_{0}^{\beta }d\tau \sum \frac{1}{2}(z_{ab}^{\mu
}+J_{ab}^{\mu })^{2}  \label{stratonovich} \\
const\cdot \exp \int_{0}^{\beta }d\tau \frac{1}{2}\sum J_{ab}^{\mu
}J_{ab}^{\mu } &=&\int Dz\exp -\int_{0}^{\beta }d\tau \sum \frac{1}{2}\left(
z_{ab}^{\mu }z_{ab}^{\mu }+2z_{ab}^{\mu }J_{ab}^{\mu }\right)   \nonumber
\end{eqnarray}
It remains to find out what energy of interaction is represented by the
expression $-\frac{1}{2}\sum J_{ab}^{\mu }J_{ab}^{\mu }$. This is done in eq(%
\ref{appendix1}) of the appendix and leads to 
\begin{eqnarray}
&&-\frac{1}{2}\sum J_{ab}^{0}J_{ab}^{0}+\overrightarrow{J}_{ab}%
\overrightarrow{J}_{ab}=\frac{1}{2}\sum \left( \psi _{a}^{+}\psi _{a}\right)
\left( \psi _{b}^{+}\psi _{b}\right) -\left( \psi _{a}^{+}\varepsilon \psi
_{a}^{+}\right) (\psi _{b}\varepsilon \psi _{b})\text{ + quadratic terms} \\
&&\text{ \ \ \ \ \ \ \ \ \ \ \ \ \ \ \ \ \ \ \ \ \ \ \ \ \ \ \ \ \ \ }%
\stackrel{N=1}{=}3n_{\uparrow }n_{\downarrow }\text{ + quadratic terms} 
\nonumber
\end{eqnarray}
where $\psi _{b}\varepsilon \psi _{b}$ means $\psi _{b\sigma }\varepsilon
_{\sigma \tau }\psi _{b\tau }$ and $\varepsilon =i\sigma ^{2}$. As
extrapolation of the Hamiltonian we choose, therefore 
\begin{equation}
H_{int}=\frac{U}{3N}\left( -\frac{1}{2}\sum J_{ab}^{\mu }J_{ab}^{\mu
}\right) =\frac{U}{6N}\sum \left( \psi _{a}^{+}\psi _{a}\right) \left( \psi
_{b}^{+}\psi _{b}\right) -\left( \psi _{a}^{+}\varepsilon \psi
_{a}^{+}\right) (\psi _{b}\varepsilon \psi _{b})  \label{Hamiltonian}
\end{equation}
This Hamiltonian collapses to the conventional expression for $N=1$ and, as
we shall see below, it scales correctly in $N$ and has a simple topological
expansion.

By appealing to the Trotter technique, the partition function of the Hubbard
model may then be represented in path integral form: 
\begin{eqnarray}
Z &=&Tr\exp -\beta \left( H_{0}+H_{int}\right)   \label{messy} \\
&=&const\int D\psi Dz\exp -\int d\tau \left[ \psi ^{\ast }(\partial
_{t}+h_{0})\psi +\frac{1}{2}\sum z_{xab}^{\mu }z_{xab}^{\mu }+\sqrt{\frac{U}{%
3N}}\sum z_{xab}^{\mu }J_{xab}^{\mu }\right]   \nonumber \\
&=&\int D\psi Dz\exp -\int d\tau \left[ \psi ^{\ast }(\partial
_{t}+h_{0})\psi +\frac{1}{2}\sum z_{xab}^{\mu }z_{xab}^{\mu }+\sqrt{\frac{U}{%
3N}}\sum \psi _{a\alpha }^{+}z_{xab}^{\mu }\sigma _{\alpha \beta }^{\mu
}\psi _{b\beta }\right]   \nonumber
\end{eqnarray}
Here $h_{0}$ is the hopping matrix of the model and we have used the fact
noted above that $z_{xab}^{\mu }$ couples only to the symmetrized part of
the current $J_{xab}^{\mu }$. A more compact and esthetically pleasing form
of the action is: 
\begin{equation}
S=\int d\tau \left( \sum \psi ^{\ast }(\partial _{t}+h_{0})\psi +\frac{1}{4}%
\sum Tr(zz^{+})+\sqrt{\frac{U}{3N}}\sum \psi ^{+}z\psi \right) \text{ with }%
z=\sum \sigma ^{\mu }z^{\mu }\text{ }  \label{action}
\end{equation}

\section{Feynman diagrams and 't Hooft's topological expansion}

From eq(\ref{messy}) and using either textbook knowledge \cite{Abrikosov} or
auxiliary currents, we find the following bare propagators of the non
interacting theory:

\begin{eqnarray}
&<&T\left\{ \psi _{a\alpha }(1)\psi _{b\beta }^{\ast }(2)\right\}
>_{0}=\delta _{\alpha \beta }\delta _{ab}\left( \frac{1}{\partial _{t}+h_{0}}%
\right) (1,2)\stackrel{\text{full propagator}}{\rightarrow }\delta _{\alpha
\beta }\delta _{ab}g(1,2)  \label{propagators} \\
&<&T\left\{ z_{a_{1}b_{1}}^{\mu _{1}}(1)z_{a_{2}b_{2}}^{\mu _{2}}(2)\right\}
>_{0}=\delta _{\mu _{1},\mu _{2}}\delta (1,2)P_{a_{1}b_{1}a_{2}b_{2}}^{\mu }%
\stackrel{\text{full propagator}}{\rightarrow }\delta _{\mu _{1},\mu
_{2}}d(1,2)P_{a_{1}b_{1}a_{2}b_{2}}^{\mu }  \nonumber \\
P_{a_{1}b_{1}a_{2}b_{2}}^{\mu } &=&\frac{1}{2}\left[ \delta
_{a_{1},a_{2}}\delta _{b_{1},b_{2}}+(-1)^{\delta _{\mu ,0}}\delta
_{a_{1},b_{2}}\delta _{b_{1},a_{2}}\right] \text{,\ \ \ \ \ \ }P^{\mu }\cdot
P^{\mu }=P^{\mu }  \nonumber
\end{eqnarray}
The projector $P^{\mu }$ reflects the difference in symmetry properties
between $z_{0}$ (antisymmetric) and $z_{i}$ (symmetric) for $i=1,2,3$. Above
we have also indicated that the tensor structure of the full propagators is
the same as that of the bare propagators, because of symmetry considerations 
\cite{comment}. There is also a vertex 
\begin{equation}
\sqrt{\frac{U}{3N}}\psi _{a\alpha }^{+}z_{xab}^{\mu }\sigma _{\alpha \beta
}^{\mu }\psi _{b\beta }  \label{vertex}
\end{equation}
where a particle hole excitation splits into its constituents. Propagators
and vertex are graphically represented in {\bf figure 1}

\begin{center}
\begin{figure}[tbp]
 \mbox{   \epsfig{file=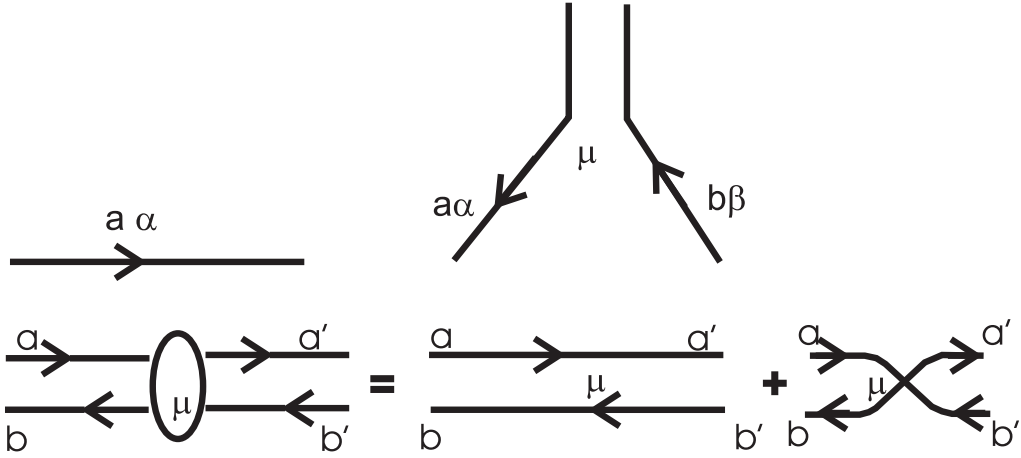,width=15cm}}
\caption{Single line electrons, double line bosons and the vertex in the $U(N,q)$
model}
\end{figure}
\end{center}

New, relative to the corresponding $SU(N)$ diagrams, is the presence of two
terms in the boson propagator. Representing fermion propagators by oriented
lines, boson double lines inherit an orientation from the electron lines
that flow through a vertex. This inherited orientation, however, is not
respected by the first term of the propagator in eq(\ref{propagators}) and a
consequence of this is illustrated in {\bf figure 2.}

\begin{center}
\begin{figure}[tbp]
 \mbox{   \epsfig{file=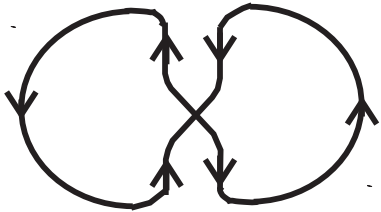,width=5cm}}
\caption{The boson propagator spoils the loop orientation}
\end{figure}
\end{center}

\noindent%
%
{}

As in $SU(N)$ Yang Mills theory \cite{Hooft} or in the $SU(N)$ Hubbard model 
\cite{DF} a diagram can be viewed as a collection of index loops that are
identified across double lines representing bosonic propagators. Each index
loop spans a two dimensional area and in this way a topological surface gets
associated with each diagram. By counting the number of vertices and index
summations from each loop, we find that each diagram carries a weight (in
powers of $N$) of 
\begin{eqnarray*}
weight &=&N^{loops-\frac{S_{0}}{2}}=N^{S_{2}-\frac{S_{0}}{2}}\stackrel{S_{1}=%
\frac{3}{2}S_{0}}{=}N^{\chi } \\
\chi  &=&S_{2}-S_{1}+S_{0}
\end{eqnarray*}
where $S_{2}$ is the number of loops or area like pieces, $S_{1}$ the number
of propagators and $S_{0}$ the number of vertices and where $S_{1}=\frac{3}{2%
}S_{0}$ relates the number of propagators and vertices. In the present $%
U(N,q)$ extrapolation, non orientable surfaces contribute and a more general
expression for the Euler characteristic must be used: 
\[
\chi =2-2h-c-o
\]
Here $h$ stands for handles, $c$ for crosscaps and $o$ for openings (see
e.g. \cite{Stillwell} for a detailed discussion of the classification of
surfaces). A ''crosscap'' is the presence, in a loop, of a pair of\ \
identified segments, that are traversed with the {\bf same} orientation. 
For large $N$, diagrams of the disk topology $h=0=c$, $o=1$
dominate and these can be easily summed (the sums of the nonleading diagrams
for $\chi =0$ and $\chi =-1$ are more complicated and will not be given
here).

\section{Coupled integral equations for the leading graphs}

Let us compute the electronic self energy to second order in perturbation
theory: 
\begin{eqnarray*}
\Sigma _{a_{1}\alpha _{1},b_{2}\beta _{2}}(1,2) &=&\frac{U}{3N}<T\left\{ %
\left[ z^{\mu }(1)\sigma ^{\mu }\psi (1)\right] _{a_{1}\alpha _{1}}\left[
\psi ^{+}(2)\sigma ^{\mu }z^{\mu }(2)\right] _{b_{2}\beta _{2}}\right\} > \\
&=&\frac{U}{3N}<T\left\{ z_{a_{1}\alpha _{1}|b_{1}\beta
_{1}}(1)z_{a_{2}\alpha _{2}|b_{2}\beta _{2}}(2)\right\} ><T\left\{ \psi
(1)_{b_{1}\beta _{1}}\psi ^{+}{}_{a_{2}\alpha _{2}}(2)\right\} >
\end{eqnarray*}
with $z_{a\alpha |b\beta }\equiv z_{ab}^{\mu }\sigma _{\alpha \beta }^{\mu }$%
. From eq(\ref{appendix2}) of the appendix we find 
\begin{eqnarray*}
\Sigma _{a_{1}\alpha _{1},b_{2}\beta _{2}}(1,2) &\equiv &\delta
_{a_{1}b_{2}}\delta _{\alpha _{1}\beta _{2}}\sigma (1,2) \\
\sigma (1,2) &=&\frac{U(2N+1)}{3N}d(1,2)g(1,2)\stackrel{N\rightarrow \infty 
}{\rightarrow }\frac{2U}{3}d(1,2)g(1,2)\text{ }
\end{eqnarray*}
Strictly speaking, this is just lowest order perturbation theory. But to
leading order in $N$ we get all the rainbow diagrams and these are summed by
rendering the outermost rainbow explicit and we do indeed obtain the last
equation with the {\bf full} propagators (to leading order in $N$), but
without vertex corrections (that are nonleading in $N$). \ We now compute
the self energy of the bosons to leading order (our notation was indicated
in eq(\ref{propagators}){\bf :} 
\begin{equation}
\Pi _{a_{1}b_{1}}^{\mu _{1}}|_{a_{2}b_{2}}^{\mu _{2}}(1,2)=\frac{U}{3N}%
<\left( \psi _{a_{1}\alpha _{1}}^{+}\sigma _{\alpha _{1}\beta _{1}}^{\mu
_{1}}\psi _{b_{1}\beta _{1}}\right) (1)\cdot \left( \psi _{a_{2}\alpha
_{2}}^{+}\sigma _{\alpha _{2}\beta _{2}}^{\mu _{2}}\psi _{b_{2}\beta
_{2}}\right) (2)>  \label{Pi}
\end{equation}
To lowest order this gives, upon using the invariant tensor properties of
the fermion propagator 
\begin{eqnarray}
\Pi _{a_{1}b_{1}}^{\mu _{1}}|_{a_{2}b_{2}}^{\mu _{2}}(1,2) &=&\pi
(1,2)\left( \text{\ }T^{\mu _{1}\mu _{2}}\right) _{a_{1}b_{1}a_{2}b_{2}}
\label{Piexplicit} \\
\left( \text{\ }T^{\mu _{1}\mu _{2}}\right) _{a_{1}b_{1}a_{2}b_{2}}
&=&\delta _{b_{1}a_{2}}\delta _{b_{2}a_{1}}\delta _{\mu _{1}\mu
_{2}}(-1)^{\delta _{\mu 0}}  \nonumber \\
\pi (1,2) &=&-\frac{2U}{3N}G(1,2)G(2,1)  \nonumber
\end{eqnarray}
It is shown in eq (\ref{appendix3}) of the appendix that the conventional
Dyson equation for the propagator leads to a corresponding scalar Dyson
equation for $d$: 
\[
d=\frac{1}{d_{0}^{-1}-\pi }
\]
The coupled integral equations 
\begin{eqnarray}
d &=&\frac{1}{d_{0}^{-1}-\pi }\text{ \ and\ \ \ }g=\frac{1}{%
g_{0}^{-1}-\sigma }  \label{selfconsistent} \\
\sigma (1,2) &=&-\frac{\delta \Phi }{\delta g(2,1)}\text{ and }\pi (1,2)=%
\frac{2}{N}\frac{\delta \Phi }{\delta d(2,1)}  \nonumber \\
\Phi  &=&-\frac{U}{3}g(1,2)d(1,2)g(2,1)  \nonumber
\end{eqnarray}
must be solved self consistently. Using the differential form of Dyson's
equations 
\begin{eqnarray}
\sigma (1,2) &=&-\frac{\delta }{\delta g(2,1)}Tr\left( \log g-g/g_{0}\right) 
\label{differentialdyson} \\
\pi (1,2) &=&-\frac{\delta }{\delta d(2,1)}Tr\left( \log d-d/d_{0}\right)  
\nonumber
\end{eqnarray}
and eq(\ref{selfconsistent}) we find that the functional

\[
\Omega =\Phi +Tr\left( \log g-g/g_{0}\right) -\frac{N}{2}Tr\left( \log
d-d/d_{0}\right) 
\]
is stationary with respect to variations in $g$ and $d$. To understand the
factor $\frac{N}{2}$, we note that the theory contains $2N$ complex fermions
and $\sim 2N^{2}$ real bosons or $\frac{N}{2}$ times as many more complex
bosons than fermions.

This leading generating functional is of the same form as a corresponding
leading functional of the $SU(N)$ theory that was written down previously 
\cite{DF} which is consistent with a general idea that different groups have
the same leading asymptotics, once their dimensions are taken large.
Differences appear, however, at the first nonleading order, because of the
nonorientable part of the boson propagator.

\section{Conclusions}

The main results of this paper are the following:

\begin{itemize}
\item  we constructed an $U(N,q)$ extrapolation of the correlation problem
that is dominated by planar diagrams at large $N$ and which permits pairing -

\item  the surfaces associated with the diagrams of this $U(N,q)$ theory now
include non orientable ones -

\item  at the leading planar level, the integral equations and generating
functional are of the same form as in the $SU(N)$ case.
\end{itemize}

It is very encouraging that the planar diagram approach to the correlation
problem permits pairing, once an appropriate extrapolation is chosen. This
is absolutely essential for its future use in investigating 
low temperature superconducting order in models of doped ceramics, organic 
conductors and other systems of interest. By using quaternion like objects,
one has kept one of the crucial features of the electrons, their spinor character.

Regarding magnetic ordering, it is unclear at present whether antiferromagnetism at
half filling will persist also for $N>1$ in this extrapolation. It is
conceivable that the magnetization $\left( \psi ^{+}L_{A}\psi \right) _{x}$,
with $L_{A}$ a basis in the space of generators of $U(N,q)$, arranges itself
to be antiparallel (in the sense of the metric $g_{AB}=TrL_{A}L_{B}^{+}$) at
neighboring points $x$ on a bipartite lattice, in a similar fashion as
discussed in \cite{Sachdev}.

It remains to find out, 
whether the small parameter 1/N introduced here is "small enough" to confer
predictive power to our approach. So far the only favorable 
indication we have is the overall agreement of the leading diagrams with 
those of FLEX plus the many interesting results obtained in this approximation 
\cite{Flexresults}. If the present approach is successful, it will sharpen the 
FLEX method and a number of tough problems in the domaine of strongly correlated 
electrons will become transparent.

\bigskip

{\bf Acknowledgments}

I am indebted to T. Dahm (Tuebingen) for continued correspondence on the
FLEX approach, to members of Laboratoire de Chimie Quantique (Toulouse) and
Theoretische Festkoerperphysik (Tuebingen) for lively discussions and useful
comments, to G. Robert and A. Zvonkin (Bordeaux) for discussions on graphs
and surfaces, to P. Sorba (Annecy) for comments on $U(N,q)$, to B.
Bonnier for encouragement and to S.Villain-Guillot for a critical reading of
the manuscript.

\section{Appendix}

\subsection{Calculation of $J_{ab}^{\protect\mu }J_{ab}^{\protect\mu }$}

We use the definition $J_{ab}^{\mu }=\frac{1}{2}\left[ \psi _{a\alpha
}^{+}\sigma _{\alpha \beta }^{\mu }\psi _{b\beta }+h.c.\right] $ with $%
\sigma ^{\mu }=\left\{ i,\overrightarrow{\sigma }\right\} $ and calculate $%
J_{ab}^{\mu }J_{ab}^{\mu }$ in two steps via $J_{ab}^{\mu }J_{ab}^{\mu
}=J_{ab}^{0}J_{ab}^{0}+\overrightarrow{J}_{ab}\overrightarrow{J}_{ab}$.
First we determine $\sum J_{ab}^{0}J_{ab}^{0}:$

\begin{eqnarray}
\sum J_{ab}^{0}J_{ab}^{0} &=&-\frac{1}{4}\sum \left[ \psi _{a\tau }^{+}\psi
_{b\tau }-\left( a\leftrightarrow b\right) \right] \left[ \psi _{a\sigma
}^{+}\psi _{b\sigma }-\left( a\leftrightarrow b\right) \right]
\label{messy1} \\
&=&\frac{1}{2}\sum -\left( \psi _{a}^{+}\psi _{b}\right) \left( \psi
_{a}^{+}\psi _{b}\right) +\left( \psi _{b}^{+}\psi _{a}\right) \left( \psi
_{a}^{+}\psi _{b}\right)  \nonumber
\end{eqnarray}
Second we determine $\sum \overrightarrow{J}_{ab}\overrightarrow{J}_{ab}:$ 
\begin{eqnarray}
\sum \overrightarrow{J}_{ab}\overrightarrow{J}_{ab} &=&\sum \frac{1}{2}\psi
_{a\alpha }^{+}\overrightarrow{\sigma }_{\alpha \beta }\psi _{b\beta }\left[
\psi _{a\gamma }^{+}\overrightarrow{\sigma }_{\gamma \delta }\psi _{b\delta
}+\left( a\leftrightarrow b\right) \right]  \label{messy2} \\
&=&\sum \frac{1}{2}\psi _{a\alpha }^{+}\psi _{b\beta }\left[ \psi _{a\gamma
}^{+}\psi _{b\delta }+\left( a\leftrightarrow b\right) \right] \left(
2\delta _{\alpha \delta }\delta _{\beta \gamma }-\delta _{\alpha \beta
}\delta _{\gamma \delta }\right)  \nonumber \\
&&\stackrel{\ast }{=}\sum -\left( \psi _{a}^{+}\psi _{b}\right) \left( \psi
_{a}^{+}\psi _{b}\right) -\left( \psi _{a}^{+}\psi _{a}\right) \left( \psi
_{b}^{+}\psi _{b}\right) -\frac{1}{2}\left( \psi _{a}^{+}\psi _{b}\right)
(\psi _{a}^{+}\psi _{b})-\frac{1}{2}\left( \psi _{a}^{+}\psi _{b}\right)
(\psi _{b}^{+}\psi _{a})  \nonumber
\end{eqnarray}
The symbol $\stackrel{\ast }{=}$ means modulo quadratic terms that
correspond to an extra chemical potential. Combining eqs(\ref{messy1},\ref
{messy2}) we obtain 
\begin{equation}
\sum J_{ab}^{0}J_{ab}^{0}+\overrightarrow{J}_{ab}\overrightarrow{J}_{ab}%
\stackrel{\ast }{=}\sum -\left( \psi _{a}^{+}\psi _{a}\right) \left( \psi
_{b}^{+}\psi _{b}\right) -2\left( \psi _{a}^{+}\psi _{b}\right) \left( \psi
_{a}^{+}\psi _{b}\right)  \label{messy3}
\end{equation}
Because there are only two spinor components we have $\left( \psi
_{a}^{+}\psi _{b}\right) \left( \psi _{a}^{+}\psi _{b}\right) =-\frac{1}{2}%
\left( \psi _{a}^{+}\varepsilon \psi _{a}^{+}\right) \left( \psi
_{b}\varepsilon \psi _{b}\right) $. Using this on eq(\ref{messy3}) we find
the required relation 
\begin{equation}
\sum J_{ab}^{0}J_{ab}^{0}+\overrightarrow{J}_{ab}\overrightarrow{J}%
_{ab}=\sum \left( \psi _{a}^{+}\varepsilon \psi _{a}^{+}\right) (\psi
_{b}\varepsilon \psi _{b})-\left( \psi _{a}^{+}\psi _{a}\right) \left( \psi
_{b}^{+}\psi _{b}\right) \text{ + quadratic terms}  \label{appendix1}
\end{equation}

\subsection{Calculation of $<T\left\{ \left[ z(1)\protect\psi (1)\right]
_{a_{1}\protect\alpha _{1}}\left[ \protect\psi ^{+}(2)z(2)\right] _{b_{2}%
\protect\beta _{2}}\right\} >$}

It is convenient to organize the algebra as follows 
\begin{equation}
<T\left\{ \left[ z(1)\psi (1)\right] _{a_{1}\alpha _{1}}\left[ \psi
^{+}(2)z(2)\right] _{b_{2}\beta _{2}}\right\} >=<T\left\{ z_{a_{1}\alpha
_{1}|b_{1}\beta _{1}}(1)z_{a_{2}\alpha _{2}|b_{2}\beta _{2}}(2)\right\}
><T\left\{ \psi (1)_{b_{1}\beta _{1}}\psi ^{+}(2)_{a_{2}\alpha _{2}}\right\}
>  \label{decomposition}
\end{equation}
$<T\left\{ z(1)z(2)\right\} >$ is linearly related to $<T\left\{
z_{a_{1}b_{1}}^{\mu _{1}}(1)z_{a_{2}b_{2}}^{\mu _{2}}(2)\right\} >$ for
which we have an expression in eq(\ref{propagators}) of the body of the
paper: 
\begin{eqnarray}
&<&T\left\{ z_{a_{1}\alpha _{1}|b_{1}\beta _{1}}(1)z_{a_{2}\alpha
_{2}|b_{2}\beta _{2}}(2)\right\} >=\sum_{\mu }\sigma _{\alpha _{1}\beta
_{1}}^{\mu }\sigma _{\alpha _{2}\beta _{2}}^{\mu }<T\left\{
z_{a_{1}b_{1}}^{\mu }(1)z_{a_{2}b_{2}}^{\mu }(2)\right\} >
\label{lotsofindices} \\
&=&\sum_{\mu }\sigma _{\alpha _{1}\beta _{1}}^{\mu }\sigma _{\alpha
_{2}\beta _{2}}^{\mu }\cdot d(1,2)\cdot \frac{1}{2}\left[ \delta
_{a_{1},a_{2}}\delta _{b_{1},b_{2}}+(-1)^{\delta _{\mu ,0}}\delta
_{a_{1},b_{2}}\delta _{b_{1},a_{2}}\right]  \nonumber \\
&=&d(1,2)\delta _{a_{1},a_{2}}\delta _{b_{1},b_{2}}\cdot \frac{1}{2}\sigma
_{\alpha _{1}\beta _{1}}^{\mu }\sigma _{\alpha _{2}\beta _{2}}^{\mu
}+d(1,2)\delta _{a_{1},b_{2}}\delta _{b_{1},a_{2}}\cdot \frac{(-1)^{\delta
_{\mu ,0}}}{2}\sigma _{\alpha _{1}\beta _{1}}^{\mu }\sigma _{\alpha
_{2}\beta _{2}}^{\mu }  \nonumber
\end{eqnarray}
with $\sigma ^{\mu }=\{i,\overrightarrow{\sigma }\}$. We use the relations $%
\frac{1}{2}\sigma _{\alpha _{1}\beta _{1}}^{\mu }\sigma _{\alpha _{2}\beta
_{2}}^{\mu }=-\varepsilon _{\alpha _{1}\alpha _{2}}\varepsilon _{\beta
_{1}\beta _{2}}$ and $\frac{1}{2}\sigma _{\alpha _{1}\beta _{1}}^{\mu
}\sigma _{\alpha _{2}\beta _{2}}^{\mu }(-1)^{\delta _{\mu ,0}}=\delta
_{\alpha _{1}\beta _{2}}\delta _{\beta _{1}\alpha _{2}}$ in eq(\ref
{lotsofindices}) to obtain: 
\[
<T\left\{ z_{a_{1}\alpha _{1}|b_{1}\beta _{1}}(1)z_{a_{2}\alpha
_{2}|b_{2}\beta _{2}}(2)\right\} >=d(1,2)\left( \delta _{a_{1},b_{2}}\delta
_{\alpha _{1}\beta _{2}}\cdot \delta _{b_{1},a_{2}}\delta _{\beta _{1}\alpha
_{2}}-\delta _{a_{1},a_{2}}\varepsilon _{\alpha _{1}\alpha _{2}}\cdot \delta
_{b_{1},b_{2}}\varepsilon _{\beta _{1}\beta _{2}}\right) 
\]
where we recognize the basic invariant tensors of the group $U(N,q)$.
Invoking the tensor character of the fermion propagator in eq(\ref
{propagators}) , the expression in eq(\ref{decomposition}) reduces to 
\begin{eqnarray}
&<&T\left\{ z_{a_{1}\alpha _{1}|b_{1}\beta _{1}}(1)z_{a_{2}\alpha
_{2}|b_{2}\beta _{2}}(2)\right\} ><T\left\{ \psi (1)_{b_{2}\beta _{1}}\psi
^{+}(2)_{a_{2}\alpha _{2}}\right\} >  \label{appendix2} \\
&=&d(1,2)\left( \delta _{a_{1},b_{2}}\delta _{\alpha _{1}\beta _{2}}\cdot
\delta _{b_{1},a_{2}}\delta _{\beta _{1}\alpha _{2}}-\delta
_{a_{1},a_{2}}\varepsilon _{\alpha _{1}\alpha _{2}}\cdot \delta
_{b_{1},b_{2}}\varepsilon _{\beta _{1}\beta _{2}}\right) \cdot g(1,2)\delta
_{b_{1}a_{2}}\delta _{\beta _{1}\alpha _{2}}  \nonumber \\
&=&d(1,2)g(1,2)\left( 2N+1\right) \delta _{a_{1},b_{2}}\delta _{\alpha
_{1}\beta _{2}}  \nonumber
\end{eqnarray}
which is the desired expression.

\subsection{Scalar form of the bosonic Dyson equation}

It remains to see how this self energy modifies the scalar coefficient $%
D(1,2)$. We must evaluate the sum 
\begin{equation}
D=D_{0}+D_{0}\Pi D_{0}+...=D_{0}\sum \left( \Pi D_{0}\right) ^{n}=D_{0}\frac{%
1}{1-\Pi D_{0}}\stackrel{??}{=}\frac{1}{D_{0}^{-1}-\Pi }  \label{tentative}
\end{equation}
$D_{0}$ is proportional to a projector\ $P$ onto matrices of definite
symmetry type, its inverse $D_{0}^{-1}$ does not exist and, at first glance,
Dyson's equation appears ill defined. In any case we need a Dyson type
equation for the scalar quantity $d$ that encapsulates the information of $D$%
. The quantities entering the series in eq(\ref{tentative}) are 
\begin{eqnarray}
D(1,2) &=&\delta _{\mu _{1},\mu _{2}}d(1,2)P^{\mu }\text{ \ \ \ \ \ \ \ \ \
\ \ \ \ ''projection''}  \label{basicdefs} \\
\Pi ^{\mu _{1},\mu _{2}} &=&\pi (1,2)\delta _{\mu _{1}\mu _{2}}(-1)^{\delta
_{\mu 0}}T^{\mu }\text{ \ \ \ ''transposition''}  \nonumber \\
\text{with} &:&\text{{}}  \nonumber \\
P^{\mu }\text{ } &=&\frac{1}{2}\left[ \delta _{a_{1},a_{2}}\delta
_{b_{1},b_{2}}+(-1)^{\delta _{\mu ,0}}\delta _{a_{1},b_{2}}\delta
_{b_{1},a_{2}}\right]  \nonumber \\
T^{\mu } &=&(-1)^{\delta _{\mu 0}}\delta _{b_{1}a_{2}}\delta _{b_{2}a_{1}}%
\text{ ''transposition''}  \nonumber
\end{eqnarray}
Our definitions were chosen so that the matrices are contracted with each
other in the same order as one writes them. The relations 
\begin{equation}
T^{\mu }P^{\mu }=P^{\mu }T^{\mu }  \label{eigenoperators}
\end{equation}
come as no surprise and they mean that the operator $\Pi ^{\mu _{1},\mu
_{2}} $ acts as a multiplication by the scalar $\pi (1,2)$. We can now sum
up the series eq(\ref{tentative}) 
\begin{eqnarray}
D &=&D_{0}+D_{0}\Pi D_{0}+...=Pd_{0}\sum_{n=0..\infty }\left( \pi
d_{0}\right) ^{n}=Pd_{0}\frac{1}{1-\pi d_{0}}=\frac{P}{d_{0}^{-1}-\pi }
\label{appendix3} \\
&\rightarrow &d^{-1}=d_{0}^{-1}-\pi  \nonumber
\end{eqnarray}
where $\left( \pi d_{0}\right) ^{n}$ is meant in the sense of convolutions
and the last equation is the desired scalar Dyson equation. \


\begin{references}
\bibitem{comment}  There is no complication due to an extra trace piece as
there was in the $SU(N)$ extrapolation
\end{references}
\end{document}